\documentclass{aa}  

\usepackage{graphicx}
\usepackage{color}
\usepackage{soul}

\DeclareUnicodeCharacter{2212}{-}

\usepackage{txfonts}
\begin{document}

   \title{Detailed chemical composition of the globular cluster Sextans A GC-1 on the outskirts of the Local Group}
 \author{A. Gvozdenko\inst{1},
          S.\ S.\ Larsen\inst{1},
          M. A. Beasley\inst{2,}\inst{3,}\inst{4},
          I. Cabrera-Ziri\inst{5},
          P. Eitner\inst{6,}\inst{7},
          G. Battaglia\inst{3,}\inst{4}, 
          \and
          R. Leaman\inst{8}
          } 

   \institute{Department of Astrophysics/IMAPP, Radboud University, PO Box 9010,
6500 GL, The Netherlands
         \and
             Centre for Astrophysics and Supercomputing, Swinburne University, John Street, Hawthorn VIC 3122, Australia
         \and
             Instituto de Astrofísica de Canarias, Calle Vía Láctea, E-38206 La Laguna, Spain
        \and    
            Departamento de Astrofísica, Universidad de La Laguna, E-38206 La Laguna, Spain
        \and
            Astronomisches Rechen-Institut, Zentrum für Astronomie der Universität Heidelberg, Mönchhofstraße 12-14, D-69120 Heidelberg, Germany
        \and
            Max-Planck Institute for Astronomy, 69117 Heidelberg, Germany
        \and
            Ruprecht-Karls-Universität, Grabengasse 1, 69117 Heidelberg, Germany
        \and
            Department of Astrophysics, University of Vienna, Türkenschanzstrasse 17, 1180 Wien, Austria
             }

   \date{Accepted  23 February 2024}

  \abstract
   {The chemical composition of globular clusters (GCs) across the Local Group provides information on chemical abundance trends. Studying GCs in isolated systems in particular provides us with important initial conditions plausibly unperturbed by mergers and tidal forces from the large Local Group spirals.}
   {We present a detailed chemical abundance analysis of Sextans A GC-1. The host galaxy, Sextans A, is a low-surface-brightness dwarf irregular galaxy located on the edge of the Local Group. We derive the dynamical mass of the GC together with the mass-to-light ratio and the abundances of the $\alpha$, Fe-peak, and heavy elements.}
   {Abundance ratios were determined from the analysis of an optical integrated-light spectrum of Sextans A GC-1, obtained with UVES on the VLT. We apply non-local thermodynamic equilibrium (NLTE) corrections to Mg, Ca, Ti, Fe, and Ni. 
   }
   {The GC appears to be younger and more metal-poor than the majority of the GCs of the Milky Way, with an age of 8.6$\pm$2.7 Gyr and $\text{[Fe/H]}=-2.14\pm0.04$ dex. The calculated dynamical mass is $M_{dyn}=(5.18 \pm1.62) \times 10^5 M_{\odot}$, which results in an atypically high value of the mass-to-light ratio, 4.35$\pm$1.40 M$_{\odot}$/L$_{V \odot}$. Sextans A GC-1 has varying $\alpha$ elements -- the Mg abundance is extremely low, Ca and Ti are solar-scaled or mildly enhanced, and Si is enhanced. The measured values are $\text{[Mg/Fe]}=-0.79\pm0.29$, $\text{[Ca/Fe]}=+0.13\pm0.07$, $\text{[Ti/Fe]}=+0.27\pm0.11$, and $\text{[Si/Fe]}=+0.62\pm0.26$, which makes the mean $\alpha$ abundance (excluding Mg) to be enhanced $\text{[<Si,Ca,Ti>/Fe]}_{\text{NLTE}}=+0.34\pm0.15$. The Fe-peak elements are consistent with scaled-solar or slightly enhanced abundances: $\text{[Cr/Fe]}=+0.31\pm0.18$, $\text{[Mn/Fe]}=+0.19\pm0.32$, $\text{[Sc/Fe]}=+0.22\pm0.22$, and $\text{[Ni/Fe]}=+0.02\pm0.12$. The heavy elements measured are Ba, Cu, Zn, and Eu. Ba and Cu have sub-solar abundance ratios ($\text{[Ba/Fe]}=-0.48\pm0.21$ and $\text{[Cu/Fe]}<-0.343$), while Zn and Eu are consistent with their upper limits being solar-scaled and enhanced, $\text{[Zn/Fe]}<+0.171$ and $\text{[Eu/Fe]}<+0.766$.
   }
   {The composition of Sextans A GC-1 resembles the overall pattern and behaviour of GCs in the Local Group. The anomalous values are the mass-to-light ratio and the depleted abundance of Mg. There is no definite explanation for such an extreme abundance value. Variations in the initial mass function or the presence of an intermediate-mass black hole might explain the high mass-to-light ratio value.}

   \titlerunning{Detailed chemical composition of the globular cluster Sextans A GC-1}
   \authorrunning{A. Gvozdenko et al.}
   
   \maketitle
%

\section{Introduction}

Observations of globular clusters (GCs) in the Milky Way (MW) and other galaxies have uncovered evidence for a ‘metallicity floor’ \citep{Harris1996, Forbes2018, Beasley2019}. An empirical minimum metallicity of [Fe/H]=-2.5 for GCs was suggested based on spectroscopic metallicities of 1928 GCs (Figure 7 in \citealt{Beasley2019}). 
Since galaxies, dwarf ones in particular, obey a well-defined mass-metallicity relation (e.g. \citealt{Kirby2013}), the existence of this metallicity floor may be related to the minimum mass that is required for GCs to survive until the present day (about $10^5 \, M_\odot$). For example, the galaxy mass-metallicity relation of \cite{Choksi2019} gives a metallicity of [Fe/H]=-2.3 at redshift $z=5$ for a mass of $10^6 \, M_\odot$, suggesting that galaxies with a significantly lower metallicity would likely not be massive enough to form GCs (see also \citealt{Kruijssen2019}). 
Recently, some GCs or tidally disrupted remains of GCs were identified to go through this metallicity floor. These include M31 GC EXT~8, [Fe/H] = -2.91 $\pm$ 0.04 \citep{Larsen2021}, and streams C-19, [Fe/H ]=-3.38 $\pm$ 0.06 \citep{Martin2022}, and the Phoenix stream, [Fe/H] = -2.7 \citep{Wan2020}. 
Another metal-poor GC that appeared to be close to the metallicity floor value is Sextans~A GC-1 \citep{Beasley2019}. It is located in an isolated dwarf irregular galaxy (dIrr).
Detailed investigation of GCs in dwarf galaxies is crucial for obtaining a complete theory of the formation and evolution of GCs. Globular clusters also provide a connection that allows one to study nearby and more distant galaxies in a consistent manner. They are used to characterise possible ancient accretion events, such as those that contributed to building the MW \citep{Bell2008, Helmi2018}. Globular clusters have been used to identify some of the past mergers by measuring their orbital and chemical properties \citep{Forbes2010, Bajkova2020}. 

\citet{Larsen2022} studied the integrated light (IL) spectra of 45 GCs in the Local Group and found for metal-poor ($\text{[Fe/H]} \lesssim -1.5$) GCs a depletion of the Mg abundance in some clusters compared to other $\alpha$ elements. The latter appears to be enhanced in all galaxies with minimal overall scatter ($<0.1$ dex). \cite{Larsen2022} found a slight preference for the GCs in dwarf galaxies to have less $\alpha$ enhancement compared to the GCs in large galaxies (by about 0.04 dex). This consistency of $\alpha$ elements suggests similar initial mass functions (IMFs) for the environments in which GCs formed. Further iron-peak and neutron-capture elements also reveal uniform abundance patterns. Some of these elements are also established to correlate with $\alpha$ elements; namely, Sc, Ni, Zn, and the light element Na.

Distinctions between the different environments in which the GCs are located are crucial for understanding the early chemical evolution of their host galaxies.
Studying isolated dIrrs, in particular, provides us with important initial conditions for tidal transformation scenarios and the opportunity to understand the evolution in low-mass halos that have not been strongly perturbed by the MW's tidal forces \citep{Kazantzidis2011, Leaman2013}.
Further, the lack of mergers in the past might manifest through a lack of bursts in the isolated objects \citep{Dohm-Palmer1997, Kennicutt2001, Forbes2022}.

The nearby Local Group galaxies such as the closest MW satellites -- the Small Magellanic Cloud and the Large Magellanic Cloud (SMC and LMC) -- provide an opportunity to study the chemical composition of stellar populations in interacting galaxies \citep{Minelli2021}. The metal-rich metallicity regime ([Fe/H] > -1 dex) of these galaxies shows large differences in the abundances in comparison with stars in the MW.
Likewise, it is important to study low-density systems. The isolated, low-mass dwarf galaxies are approximations in nature to ‘closed’ models of chemical enrichment\footnote{'approximations' because they may still have experienced feed-back driven outflows in the past} as they supposedly have neither undergone any strong tidal interactions nor experienced ram pressure stripping. 

One of the excellent targets in the nearby galaxies is Sextans~A GC-1, mentioned above. This GC is located in a low-surface-brightness dIrr, Sextans~A, which has a peculiar diamond shape. This galaxy is seen nearly face-on with an inclination of i$\sim 36^o$ \citep{Skillman1988}. 
\cite{Dolphin2003} used Cepheids, red clump stars, and the tip of the red giant branch (RGB) to measure the distance, which was derived to be 1.32$\pm$0.04 Mpc. Later this value was updated by \cite{Bellazzini2014} to 1.42$\pm$0.08 Mpc. This value puts Sextans~A right at the edge of the Local Group.

Studies of existing stellar populations in Sextans~A show that the star formation (SF) was not continuous but bursty. It had a significant SF in ancient times, >5-10 Gyr ago, with a break at intermediate ages (1 - 5 Gyr ago), and a bursty episode that started between 1 and 2.5 Gyr ago \citep{Mateo1998}. It has continuously formed stars since then. In particular, the SF rate over the past 0.06 Gyr is a factor of 20 larger than the average over the whole lifetime of the galaxy. A small number of stars older than 2.5 Gyr was found; however, there was no evidence of other older bursts in the past besides the ancient epoch of SF at $>$5-10 Gyr\citep{vanDyk1998, Dohm-Palmer1997, Dolphin2003_3, Garcia2019}.

Sextans~A is quite isolated, as the nearest galaxy to it is 300 kpc away (Sextans~B). Therefore, it is suggested that the SF is mainly due to intrinsic processes and is unrelated to tidal action from a nearby neighbour. Over this distance, the dynamical effects are expected to be minimal \citep{vanDyk1998}. This means that the Sextans~A and its GCs should have no input from extragalactic material as it is a low-density system.

It is fascinating to see what the detailed chemical composition can tell us about the outskirts of the Local Group in isolated systems such as Sextans~A GC-1.
\cite{Beasley2019} studied this cluster using the OSIRIS integrated spectrum and Du Pont (100-inch telescope at the Las Campanas Observatory) V and I band imaging.
Using photometry, these authors measured a circularised half-light radius to be 1.10 $\pm$ 0.03 arcsec, which is covered well by the OSIRIS slit width of 1.20 arcsec. 
The cluster is quite elliptical, $\epsilon$=0.12 $\pm$ 0.01, and at the distance D=1.42 $\pm$ 0.08 Mpc \citep{Beasley2019} the half-light radius corresponds to $R_h=7.6 \pm 0.2$ pc \cite{Beasley2019}. 
The cluster is thus quite large and close in size to the median of GCs in the outer halo of M31 and bigger than the median size of GCs in the MW and nearby dIrrs \citep{Harris1996, Georgiev2009, Huxor2014}. 
\citet{Beasley2019} derived the first estimates of age (8.6 $\pm$ 2.7 Gyr), the iron abundance [Fe/H]=-2.38$\pm$0.29), and the heliocentric radial velocity ($ \nu _{helio}$=305$\pm$15 km~s$^{-1}$); while the heliocentric velocity of the system Sextans~A is $324\pm2$ km~s$^{−1}$ from HI \citep{Koribalski2004}. 

In this study, we revisit Sextans~A GC-1 to examine its low iron abundance and further derive the detailed chemical composition of this GC. Additionally, using the high-resolution IL spectrum we derive the velocity dispersion, which allows us to derive the dynamical mass and the mass-to-light ratio (M/L). 
Sections are as follows: Section\,\ref{data} presents the observations we used, which are described together with the data reduction. The main stages of the analysis method are explained in Section\,\ref{methods}, including the secondary method of defining the estimated age and metallicity (Section\,\ref{analysis}). The results are given in Section\,\ref{results}. We discuss and compare our findings to other Local Group GCs in Section\,\ref{discussion}. We offer a discussion of the dynamical mass and the mass-to-light ratio in Section\,\ref{Mass and ratio}, while the age metallicity relation is discussed in Section\,\ref{age Z}.
In Section\,\ref{conclusions}, we present our main conclusions.

\begin{table*}
\caption{Literature and derived basic parameters of Sextans~A and its GC-1.}           
\label{galaxy_param} 
\centering 
\begin{tabular}{l c l}   
\hline
\hline  
Quantity & Observed value & References \\ 
\hline    
 & \textit{Sextans~A} & \\
\hline 
    D & 1.42 $\pm$ 0.08 Mpc & \cite{Bellazzini2014} \\ 
   SFR (lifetime) & 600 $\pm$ 300 \(M\odot\) Myr$^{-1}$ kpc$^{-2}$ & \cite{Dolphin2003}   \\
   SFR (last 2.5 Gyr) & 1600 $\pm$ 500 \(M\odot\) Myr$^{-1}$ kpc$^{-2}$& \cite{Dolphin2003}   \\
     
   \text{[M/H]} & $\approx$ -1.4 & \cite{Dolphin2003}  \\

\hline
 & \textit{Sextans~A GC-1} & \\
 \hline
 R.A. & 10:10:43.80 & \\
 Dec & - 04:43:28.8 & \\
 V & 18.04 & \cite{Beasley2019} \\  
 V-I & 0.98 & \cite{Beasley2019} \\  
 RV & 340.4$\pm$0.6 km s$^{-1}$ & this work \\  
 Age & $\sim$9 Gyr & this work\\
    & 8.6$\pm$2.7 Gyr & \cite{Beasley2019} \\
 \text{[Fe/H]} & - 2.14$\pm$0.04 & this work\\
 M$_{dyn}$ & (5.18 $\pm1.62) \times 10^5 M_{\odot}$ & this work\\
 M/L & 4.35$\pm$1.40 M$_{\odot}$/L$_{V \odot}$ & this work\\
  
\hline 
\end{tabular}
\tablefoot{D is the distance to the galaxy, SFR is the SF rate, M$_{dyn}$ is the dynamical or virial mass, and M/L is the mass-to-light ratio}
\end{table*}

\section{Data}
\label{data}

\begin{figure*}
\centering
\sidecaption
\includegraphics[width=12cm]{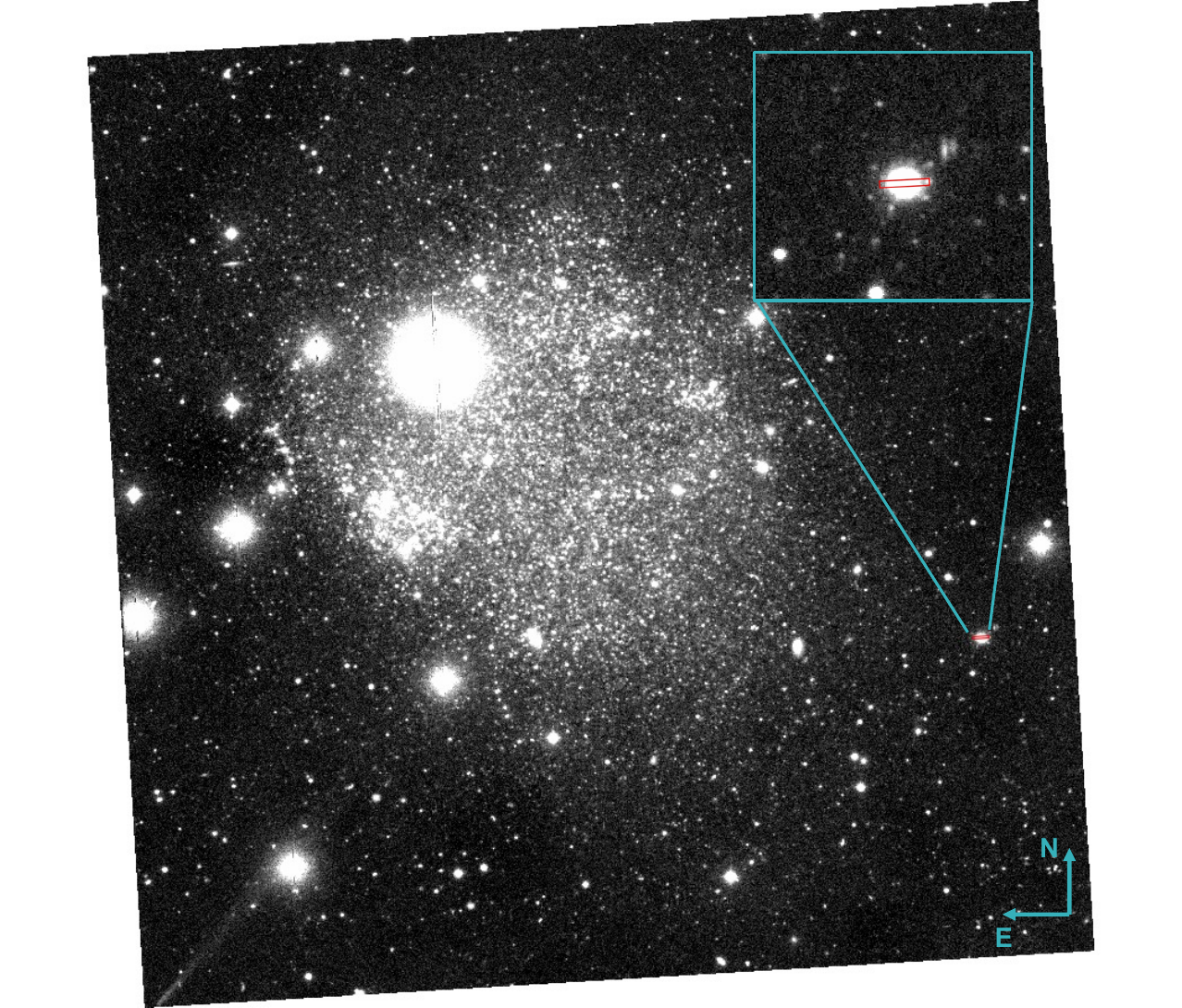} 
\caption{Image of the Sextans A galaxy using du Pont 100 inch. The slit drawn on top of Sextans A GC-1 indicates the location of the UVES slit.} 
\label{target_img}%
\end{figure*}

The basic parameters of Sextans~A and the Sextans~A GC-1 are listed in Table\,\ref{galaxy_param}, while the image of the target is shown in Figure\,\ref{target_img} from the Du Pont telescope.

Sextans~A GC-1 was observed with UVES on the ESO Very Large Telescope \citep{Dekker2000} in 2021 (14-15 April, 18-19 April, and 6-7 June).
The observation of the target was performed using the red arm and executed in clear conditions and
a seeing of $1\farcs5$, thin sky transparency, and airmass ranging between 1.06 and 1.39.

The target's magnitude is V=18.04 with five hours of observational time, resulting in a signal to noise (S/N) per \AA\ of 160 at 6000 \AA. 
The data were reduced (flat-fielding, wavelength calibration, and flux calibration) using EsoReflex (esoreflex-2.11.5) for UVES (uves-6.1.8). 
The final spectrum covers the region of 4167-6205 \AA. The resolution-slit product of the red arm on UVES is 38,700 and the slit width used was 1.20 arcsec, yielding a spectral resolving power of 33,000 \citep{Dekker2000}.

During the analysis to assess the age estimate, a method from \cite{Cabrera-Ziri2022} was applied.
For this method it is favourable for a spectrum to contain Ca H+K lines (see Section\,\ref{analysis}); therefore, an IL spectrum obtained with OSIRIS, GTC (La Palma) was a better option \citep{Cepa2010}.
The observations were performed on March 5 2016 and the data was obtained for and adopted in \cite{Beasley2019}. 
The slit width of 1 arcsecond and 2000B grism were used, and the integrated time was 600 s.
The data were reduced using IRAF and Python scripts combined. The covered wavelength of the final spectrum is 3947 -- 5693 \AA\, the resolution (FWHM) is 3.0$\pm$0.2 \AA\, and the median S/N is 27 \AA$^{-1}$.

\section{Analysis} 

\label{methods}
\subsection{Models and stellar parameters}
The analysis in this study was performed in a similar manner to \cite{Larsen2022}, in order for the results to be consistent and comparable. This section covers only the basic overview of the method, while the detailed analysis of high-resolution IL data for star clusters is described in \cite{Larsen2022}.

The analysis was performed with a Python3 package called ISPy3 \citep{Larsen2020}. 
The method uses full spectral synthesis, which automatically accounts for line blending, and one can access a wide variety of elements: $\alpha$ abundances (Mg, Si, Ca, and Ti), Fe-peak elements (Cr, Mn, Fe, Ni, and Sc), heavy elements (Cu, Zn, Zr, Ba, and Eu), and lighter elements such as Na. 
At first, the spectrum was split up into a number of segments for a pre-analysis step (approximately 100 or 200 Å each); this allowed us to estimate the broadening per segment, which was then used to calculate the mean and standard deviation of the broadening. 
Subsequently, only regions with lines from specific elements were fitted, and hence spectral windows were provided for each element where their lines are present.
Spectral windows were prepared based on the \cite{Wallace2000} spectrum of Arcturus and the solar spectrum (\citealt{Kurucz1984}; 2005 version).
The spectral windows are the same as in \cite{Larsen2022} with a few exceptions.
For some elements, the spectral windows had to be adjusted due to the wavelength coverage of the current data set. A couple of Fe spectral windows were shortened (4952.0-4962.0 \AA\ and 5610.0-5630.0 \AA) and one window was excluded completely (5008.0 -5017.0 \AA). 
The Mg b triplet (5523.0-5531.0 \AA) was excluded as it falls on the edge of the UVES coverage. 
However, for some of the used spectral windows, the analysis could not find a solution, typically because the spectral features in these windows were too weak to constrain the corresponding abundance at the S/N of our data. The full list of the successfully used spectral windows is given in Table\,\ref{tab:spec_win_abun}. 
{The table also lists the S/N per \AA\ for each spectral window. These values were computed from the S/N per pixel for reduced data:
\begin{equation}
    S/N_{\AA} = S/N_{pix} \times \sqrt{a}
,\end{equation}
where \textit{a} is the dimensionless number of pixels in one \AA: 
1 \AA\ $\div$ 0.026 \AA\ $\approx$ 38.

As a starting point for the analysis, we based the spectral modelling on an isochrone with a metallicity similar to that derived from the previous spectroscopic analysis, [Fe/H]=-2.4.
In subsequent iterations, the choice of isochrone was updated for consistency with our spectroscopic analysis.
In our analysis, the stellar populations were based on the theoretical isochrones of the Dartmouth Stellar Evolution Program (DSEP) \citep{Dotter2007}. These isochrones cover stellar populations up to the tip of the RGB evolutionary stage. Therefore, to account for the horizontal branch (HB) stars, photometry of MW GCs was used to complement the DSEP isochrones. 
The HBs were taken from Advanced Camera for Surveys photometry in \cite{Sarajedini2007} and merged with the isochrones in order to produce a complete Hertzsprung-Russell Diagram (HRD) (see \cite{Larsen2022} for details).
While other isochrone libraries exist that have more complete coverage of the HRD (including the HB and AGB), we continued using the combination of DSEP isochrones and empirical HB data for consistency with the previous work by \cite{Larsen2022}.
All these different stellar evolutionary stages from the entire HRD with a combination of HB and DSEP were organised into a set of 98 bins, each with a number of parameters defining the spectral synthesis. The parameters were stellar mass, effective temperature ($T_\mathrm{eff}$), the logarithm of surface gravity ($\log(g)$), the radius of the star, the weight of the bin (i.e. the number of stars in the mass range covered by the bin), the logarithm of microturbulent velocity, and the atmospheric model and spectral synthesis codes used to produce the synthetic spectra. The two different combinations were ATLAS12+\texttt{SYNTHE} for stars with $T_\mathrm{eff}>4000$~K \citep{Kurucz1970} and MARCS+\texttt{Turbospectrum} for the cool stars ($T_\mathrm{eff}<4000$~K) \citep{Gustafsson2008, Plez2012}. 
When co-adding the model spectra for each bin to produce an integrated-light spectrum, weights were assigned to each bin, assuming a stellar mass function of the form $dN/dM\propto M^{^\alpha}$. A value for the exponent of $\alpha=1$ was assumed in order to account for the preferential loss of low-mass stars due to dynamical evolution \citep{Sollima2017, Larsen2022}. A lower mass limit of $0.15 M_\odot$ was assumed.
A microturbulent velocity, v$_{\text{turb}}$, for each bin, is described through a linear function of logarithmic surface gravity, log($g$):
\begin{equation}
     v_{\text{turb}} = 1.88 - 0.23 \text{log}(\textit{g}))\text{km~s}^{-1}
,\end{equation}
while for the HB stars the value of the microturbulent velocity was set to v$_{\text{turb}}$=1.8 $\text{km~s}^{-1}$ \citep{Pilachowski1996, Larsen2022}. 
The abundances used in the modelling were then iteratively adjusted until the best fit to the observed spectra was obtained. The best fit was estimated based on the $\chi^2$ of the difference between the model spectra and the observed spectra, once the model spectra had been scaled to the observed spectrum by means of spline or polynomial fitting functions (to match the continuum levels).
The errors for derived abundance values (listed in Table\,\ref{tab:spec_win_abun}) were assessed when the $\chi^2$ value 
increased by one, compared to the best-fit value, while varying the abundances.
}

\subsection{Analysis of the Sextans~A GC-1}
\label{analysis}

In order to define the HRD, an assumption about age and metallicity is required. 
As discussed above, we used an isochrone with a metallicity that self-consistently matched that derived from our spectral analysis. However, the age was not well constrained by our analysis of the UVES spectra.
To verify the values obtained in \cite{Beasley2019}, we used the method from \cite{Cabrera-Ziri2022}.
This approach uses the integrated light spectrum to infer the HB properties, age, metallicity, and individual element abundance. For this method, we used the OSIRIS data due to its wavelength coverage.
The OSIRIS spectrum has a blue spectral coverage that includes the \ion{Ca}{ii} H and K lines, which  are important for constraining the morphology of the HB.
The derived estimate of the age was at least $\sim$ 9 Gyr, which is in agreement with \cite{Beasley2019}, and this analysis also confirmed that GC-1 is metal-poor, [Fe/H] $\sim$ -2.00, and that it probably hosts a prominent hot HB population. 
That said, we should note that although these independent estimates seem consistent with the literature values, the metallicity of this target is outside the regime in which the models used in \cite{Cabrera-Ziri2022} were designed to operate (i.e. $\sim$ -1.5 < [Fe/H] < 0.3, see \citealt{Conroy2018}). Hence, we should sound a note of caution regarding the accuracy of these results.

After a number of iterations including isochrones with various metallicity values, including -2.40 from \cite{Beasley2019} and -2.00 from the above analysis of \cite{Cabrera-Ziri2022},
the final HRD used corresponded to an $\alpha$-enhanced isochrone with an age of 9 Gyr and a metallicity of $-2.20$. The appointed HB was from the Galactic GC NGC~7099, which has a metallicity of [Fe/H]=$-2.24\pm0.02$, similar to that of GC-1 \citep{Larsen2022}. 

The radial velocity derived from the UVES spectrum is 340.4$\pm$0.6 km s$^{-1}$. It was obtained with an iterative approach in which the radial velocity was determined in a number of the wavelength intervals, from which the mean and the uncertainty on the mean, using the standard deviation, were computed.
The derived radial velocity with this method is higher than the earlier determined value of 305$\pm$15 km s$^{-1}$ in \cite{Beasley2019}. Sextans A GC-1 then has a radial velocity offset of about 16 km~s$^{-1}$ with respect to the Sextans A (\ion{H}{i}) systemic velocity (324 $\pm$ 2 km s$^{-1}$) \citep{Koribalski2004}.
The broadening velocity required to fit the model spectra to the UVES observations
is $\sigma$=6.67$\pm$0.69 km s$^{-1}$. Once the UVES instrumental broadening of 3.9 km s$^{-1}$ is subtracted in quadrature \citep{Dekker2000}, the velocity broadening of GC-1 itself becomes 5.41$\pm$0.77 km s$^{-1}$.

\section{Results}

\label{results}

We have derived the values of iron (Fe), $\alpha$- (Mg, Ca, Ti, Si), Fe-peak (Cr, Mn, Sc, Ni), and heavy (Ba, Cu, Zn, Eu) element abundances. We also give the non-local thermodynamic equilibrium (NLTE) corrected values for elements for which these have been determined (Mg, Ca, Ti, Mn, Fe, Ni, and Ba) \citep{Eitner2019, Larsen2022}. 
The rest of the elements (Si, Cr, Sc, Cu, Zn, and Eu) are unaltered and only the local thermodynamic equilibrium (LTE) values are given.

\begin{table*}
\caption{Derived abundances.}
\label{tab:abundances}
\centering
\begin{tabular}{cccccc|c}
\hline
\hline
Abundance &  & &  & & &
\\
ratio & LTE & S$_{\text{X}}$ & NLTE & S$_{\text{X}}$ & $\sigma_{\langle \text{X} \rangle}$ & $\Delta\sigma=\pm0.7$ km s$^{-1}$ 
\\ 
\hline
[Fe/H] & - 2.17 & 0.04 &  -2.14 &  0.04 & 0.02 & $\pm$0.05
\\
{[Mg/Fe]} & - 0.78 & 0.09  & - 0.79  & 0.10 & 0.29 & $\mp$0.02
\\
{[Ca/Fe]} & +0.13 & 0.07 &  +0.13  & 0.07 & 0.05 & +0.01
\\
{[Ti/Fe]} & +0.11 & 0.12 &  +0.27  & 0.11 & 0.07 & 0.00
\\
{[Si/Fe]} & +0.62 & 0.26 & - & -& 0.26  & +0.04
\\
{[$\alpha$/Fe]} & +0.29 & 0.15 &  +0.34  & 0.15 & 0.13 & +0.10
\\
{[Cr/Fe]} & +0.31 & 0.18 &  - &- & 0.07 & +0.01
\\
{[Mn/Fe]} & - 0.01 & 0.31 &  +0.19  & 0.32 & 0.18 & +0.04
\\
{[Sc/Fe]} & +0.22 & 0.22 &   -& -& 0.11 & $\mp$0.01
\\
{[Ba/Fe]} & - 0.30 & 0.22 &  -0.38  & 0.22 & 0.12 & $\pm$ 0.05
\\
{[Ni/Fe]} & - 0.14 & 0.14 &  +0.02  & 0.12 & 0.10&  +0.01
\\
{[Cu/Fe]} &  <-0.355 & &    & &  & +0.02
\\
{[Zn/Fe]} & <+0.143 & &    & &  & $\pm$0.02
\\
{[Eu/Fe]} & <+0.766 & &    & & & $\mp$ 0.11
\\
\hline  
\end{tabular}
\tablefoot{$\Delta\sigma$ is the forced change in velocity dispersion to show the effect on each of the element abundances.}
\end{table*}

Each element had one or more spectral windows fitted; table\,\ref{tab:spec_win_abun} provides the full list of wavelength ranges, the derived values (LTE and NLTE), and their uncertainties from ISPy3. These values were used to calculate the final value of each element. The elements with a single spectral window (e.g. Si, Cu, Zn, and Eu) have values directly from ISPy3 output. 

For some of the lines, only upper limits were found in some cases (i.e. lines were too weak). 
In these cases, while the fit converged towards a best-fit value, no lower bound on the uncertainty range was found (i.e. the data are consistent with the complete absence of the line), and hence only a positive error bar is defined.
The values and uncertainties listed in Table\,\ref{tab:abundances} are the weighted average of the values derived for the spectral windows of every individual element:

\begin{equation}
\langle [\text{X/Fe}] \rangle=\frac{\sum w_i[\text{X/Fe}]_i}{\sum w_i}
\label{eqn:mean abundance}
.\end{equation}

There are two different uncertainties listed for the final abundance ratio values in the Table\,\ref{tab:abundances}. The first one is $\text{S}_{X}$ − an estimate of the standard error on the mean abundance, which was derived using the weighted standard deviation ($\text{SD}_{X,w}$):

\begin{equation}
\text{SD}_{X,w}= \left[ \frac{N}{N-1} \frac{\sum (\text{X/Fe}_i - \langle \text{X/Fe} \rangle )^2 w_i}{\sum w_i} \right]^{1/2}
\end{equation}
\begin{equation}
S_{\text{X}}=SD_{X,w}/ \sqrt{N}
.\end{equation} 

The weights were defined as

\begin{equation}
w_i=\frac{1}{\sigma_i^2+\sigma_0^2}
,\end{equation}

where $\sigma_0 = 0.05$ dex was used to account for non-random uncertainties on the derived individual window abundances. Such non-random uncertainties refer to a typical scatter in the abundances between spectral windows that is seen even when spectra of very high S/N are analysed \citep{Larsen2022}.

Another way of estimating the uncertainties on the mean value is to propagate the individual measurement uncertainties, according to the usual formula:

\begin{equation}
\sigma_{\langle \text{X} \rangle}=\left( \sum w_i \right)^{-1/2}
.\end{equation}
Both estimates of the uncertainties are listed in Table\,\ref{tab:abundances}. While S$_X$ can vary between the LTE and NLTE values depending on how the individual measurements change, $\sigma_{\langle X\rangle}$ stays the same and is therefore only listed per abundance ratio.
In the following, we generally use the larger of the two uncertainties.

To assess how the uncertainty on the velocity dispersion affects our abundance measurements, we performed a test in which we varied the velocity dispersion within the uncertainty range and then repeated the abundance measurements (see the right column in Table\,\ref{tab:abundances}). The effect is small for the examined range ($\pm$0.7 km s$^{-1}$) and it can be said that velocity broadening is not a strong contribution to the systematic uncertainty.
The age difference was explored in previous literature; for example \cite{Larsen2014}, where the authors find that the age (within several Gyr) of the input isochrone has only a small effect on the results ($<0.1$~dex for [Fe/H] and $<0.05$ for abundance ratios [X/Fe].

We used 36 windows containing lines of \ion{Fe}{i} and \ion{Fe}{ii} (see Table.\ref{tab:spec_win_abun} for the list of spectral windows fitted). Sextans~A GC-1 was found to be quite metal-poor, $\text{[Fe/H]}=-2.14\pm0.04$.
The abundance ratios of individual $\alpha$ elements differ strongly. In particular, Mg is depleted with respect to the solar value, $\text{[Mg/Fe]}=-0.79\pm0.28$, 
while the values for Ca, Ti, and Si are all consistent with super-solar [$\alpha\text{/Fe]}$ ratios ($\text{[Ca/Fe]}=+0.13\pm0.07$, $\text{[Ti/Fe]}=+0.27\pm0.11$, and $\text{[Si/Fe]}=+0.62\pm0.26$).
The overall $\alpha$ abundance, [$\alpha\text{/Fe]}=+0.34\pm0.15$, was calculated using only the abundance ratios of [Ca/Fe], [Ti/Fe], and [Si/Fe], due to the extremely depleted [Mg/Fe] value (see Section\,\ref{Subsect:mg} for further discussion).
Iron-peak elements derived include Sc, Cr, Mn, and Ni. 
The Fe-peak elements are consistent with scaled-solar or somewhat elevated abundance ratios: $\text{[Sc/Fe]}=+0.22\pm0.22$, $\text{[Cr/Fe]}=+0.31\pm0.18$,  $\text{[Mn/Fe]}=+0.19\pm0.32$, and  $\text{[Ni/Fe]}=+0.02\pm0.12$.
The heavy elements derived for this GC are Ba, Cu, Zn, and Eu. Ba was derived using three spectral windows. The derived value for the Ba abundance ratio is sub-solar, $\text{[Ba/Fe]}=-0.38\pm0.22$. The other three elements had a single spectral window and only an upper limit of the abundance was found. 
The upper limit of Cu is also sub-solar ($\text{[Cu/Fe]}<-0.355$), and for Zn the upper limit is consistent with a scaled-solar composition ($\text{[Zn/Fe]}<+0.143$), whereas the upper limit of Eu allows a significantly super-solar abundance ratio, $\text{[Eu/Fe]}<+0.766$.

\begin{figure*}
\centering
\sidecaption
\includegraphics[width=12cm]{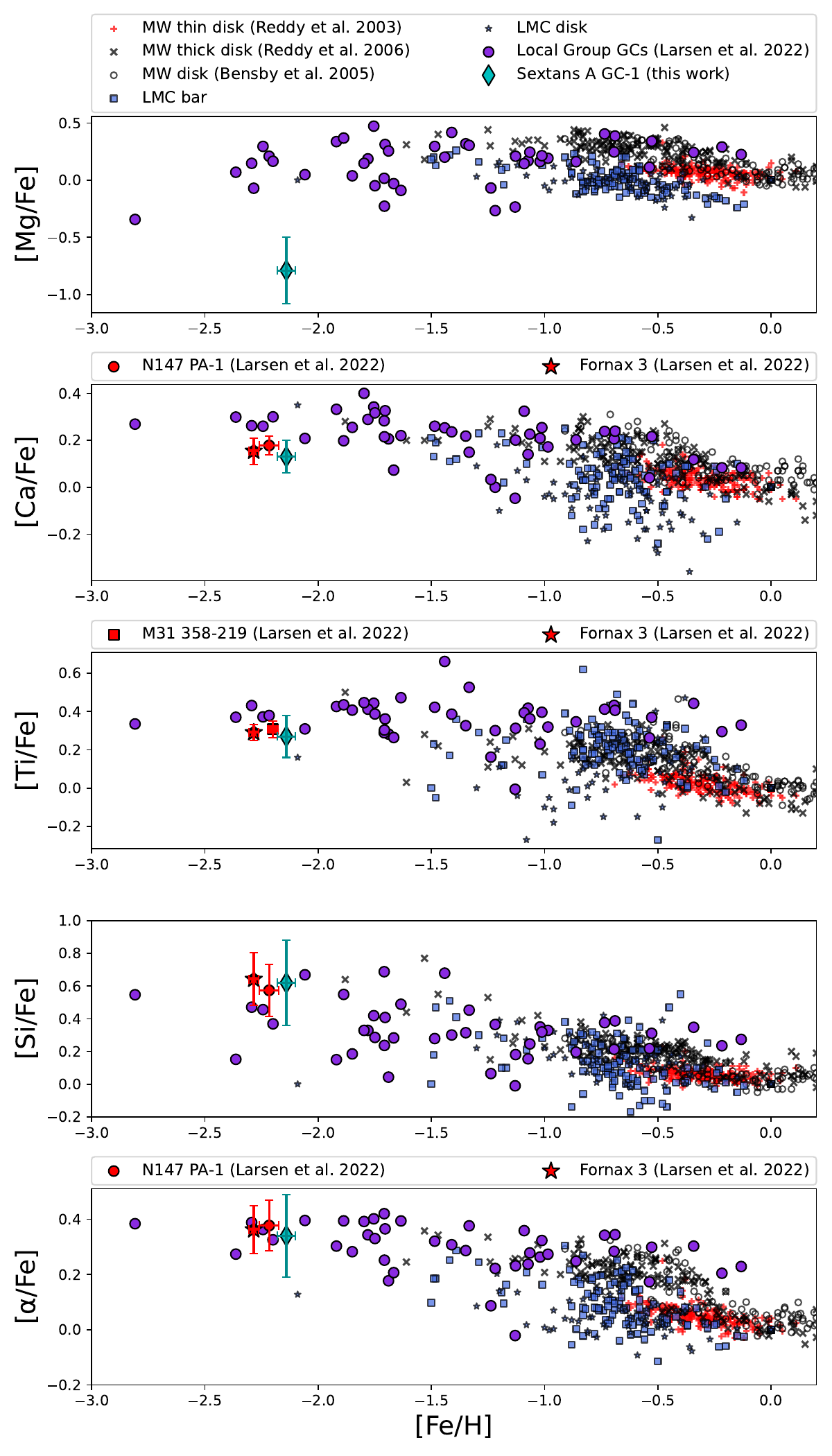}
\caption{Individual $\alpha$ elements plotted against iron abundance, [Fe/H]. Top: [Mg/Fe] vs [Fe/H]. Second row: [Ca/Fe] vs [Fe/H]. Third row: [Ti/Fe] vs [Fe/H]. Fourth row: [Si/Fe] vs [Fe/H]. Bottom: $\text{[<Si,Ca,Ti>/Fe]}$ vs [Fe/H]. Turquoise rhombus refers to this work. Purple circles show the NLTE abundance of the GCs in the Local Group from \cite{Larsen2022};
some of these Local Group GCs are highlighted in red symbols and introduced in the legends accordingly, e.g. N147 PA-1 is a red circle, Fornax~3 is a red star, M31 358-219 is a red square, 
and red crosses and black Xs present MW disc abundances from \cite{Reddy2003, Reddy2006}, respectively. 
Black open circles are the MW disc abundances from \cite{Bensby2005}.
Blue squares and stars belong to LMC bar and inner disc abundances presented by \citealt{VanderSwaelmen2013}).
} 
\label{Fig_alpha_elemVSfe}%
\end{figure*}

\begin{figure}
\centering
\includegraphics[width=0.49\textwidth]{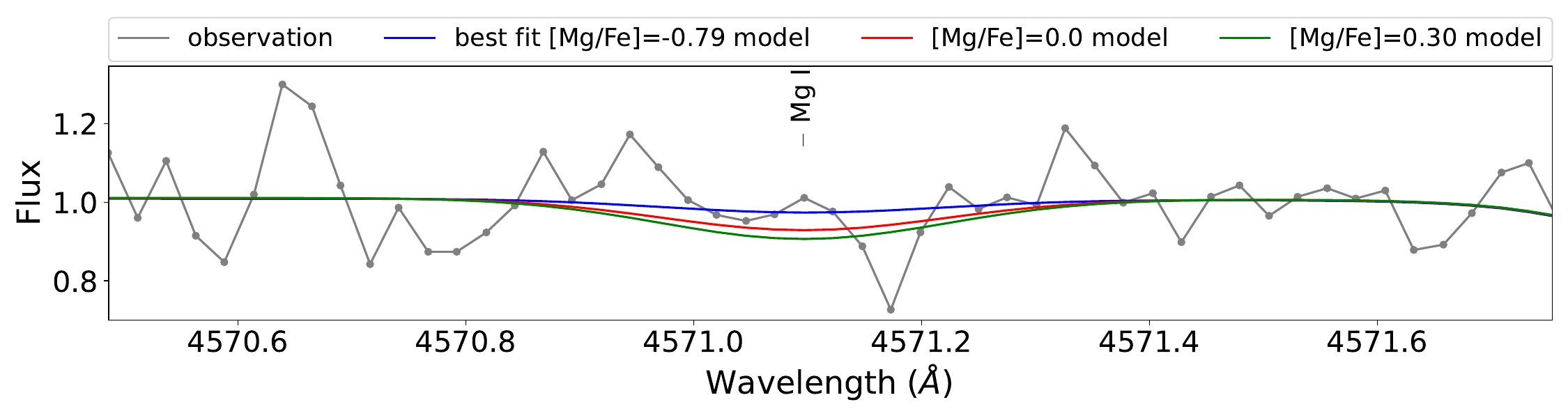}
\includegraphics[width=0.49\textwidth]{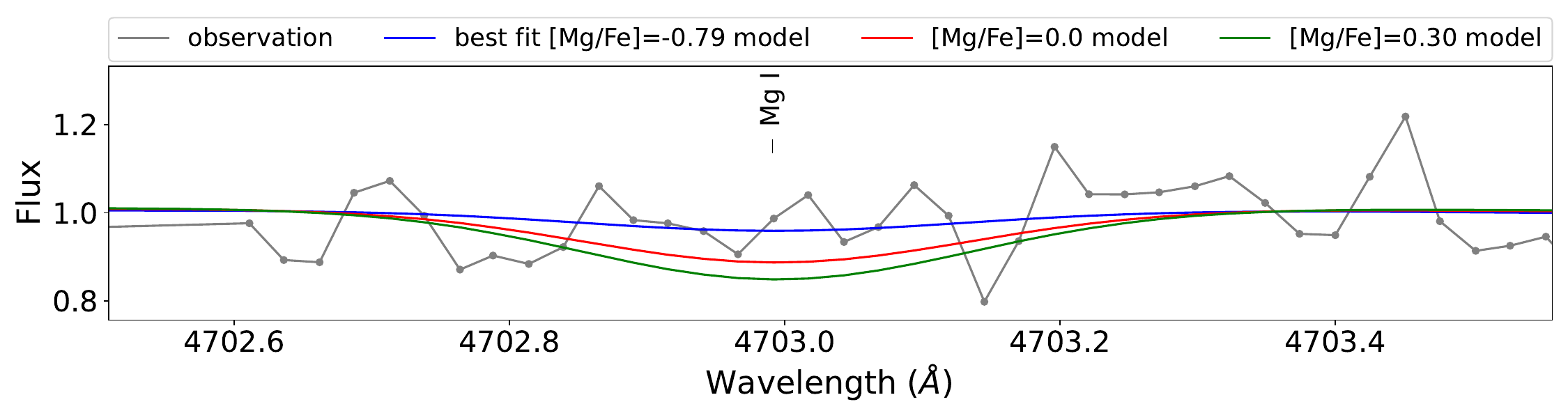}
\caption{Variation in [Mg/Fe] test. Top: Spectral window of the Mg I line at 4571 \AA. Bottom: Spectral window of the Mg I line at 4703 \AA.
The best-fit model is in blue, the solar value of [Mg/Fe] is in red, the enhanced [Mg/Fe] abundance is in green, and the observed spectrum is in grey.} 
\label{Fig_Mg_test}%
\end{figure}

\begin{figure*}
\centering
\sidecaption
\includegraphics[width=12cm]{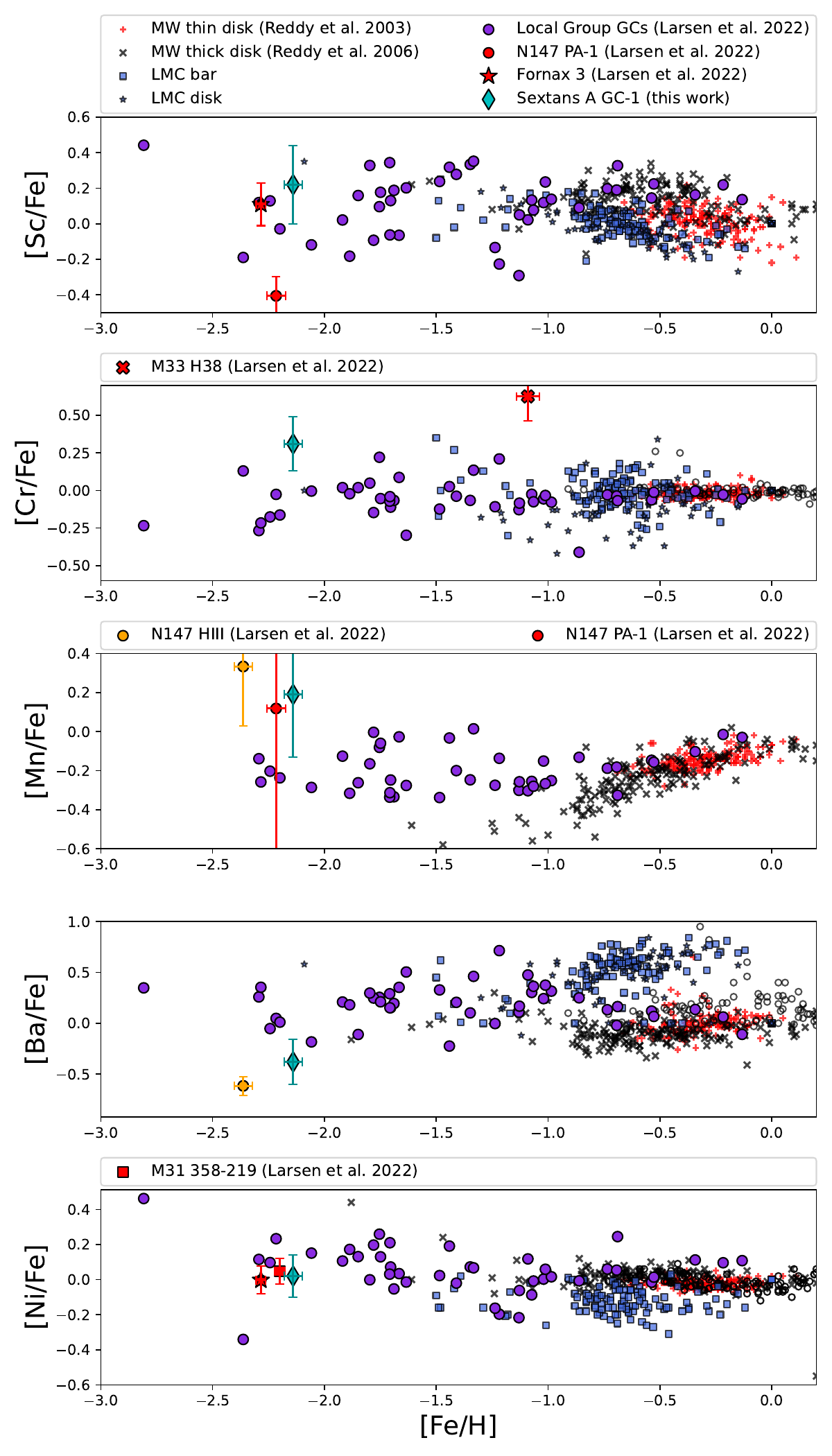}
\caption{Fe-peak and heavy elements plotted against iron abundance, [Fe/H]. Top: [Sc/Fe] vs [Fe/H]. Second row: [Cr/Fe] vs [Fe/H]. Third row: [Mn/Fe] vs [Fe/H]. Fourth row: [Ba/Fe] vs [Fe/H]. Bottom: [Ni/Fe] vs [Fe/H].  Symbols are the same as in Figure\,\ref{Fig_alpha_elemVSfe}.
} 
\label{Fig_other_elemVSfe}%
\end{figure*}

\begin{figure*}
\centering
\sidecaption
\includegraphics[width=12cm]{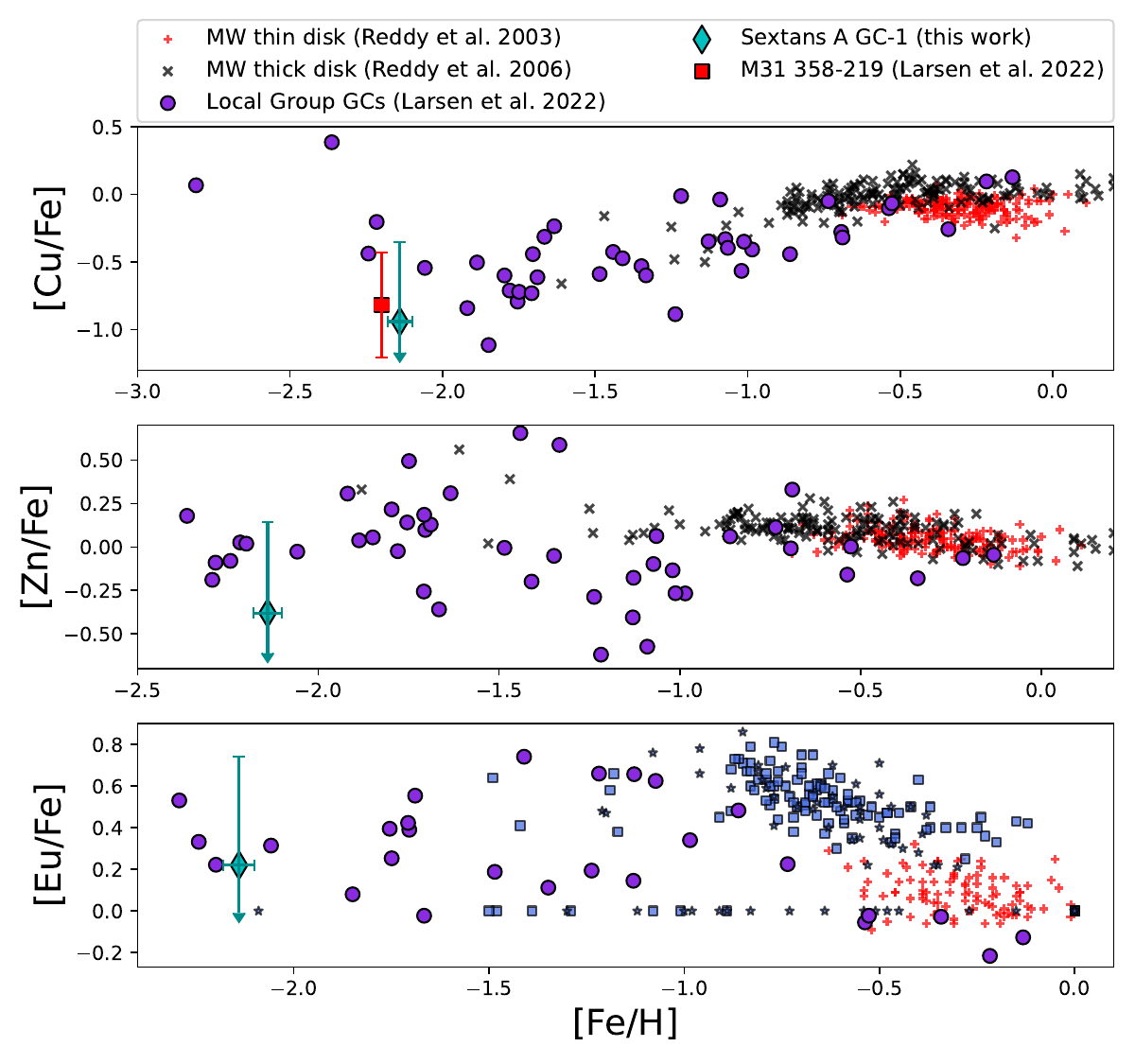}
\caption{Cu, Zn, and Eu abundance plotted against iron abundance, [Fe/H]. Top: [Cu/Fe] vs [Fe/H]. Middle: [Zn/Fe] vs [Fe/H]. Bottom: [Eu/Fe] vs [Fe/H]. The empty symbol represents the upper limit derived for the element. The line represents all the possible values for this element. Symbols are the same as in Figure\,\ref{Fig_alpha_elemVSfe}.
} 
\label{Fig_range_elemVSfe}%
\end{figure*}

The data allowed us to measure the virial mass and the mass-to-light ratio using the line-of-sight velocity broadening of $\sigma=5.41 \pm 0.77$ km~s$^{-1}$ calculated in Section\,\ref{analysis}.
This was then used to calculate the virial mass,
\begin{equation}
M_{dyn}=10 \frac{\sigma^2 r_h}{G}=(5.18 \pm1.62) \times 10^5 M_{\odot},
\end{equation}
where the half-light radius is taken from \cite{Beasley2019}, 7.6$\pm$0.2 pc.
The same authors obtained the absolute magnitude, M$_{V,0}$=-7.85, from the integrated magnitude, V=18.04. This results in an unexpectedly high value for the mass-to-light ratio of M$_{dyn}$/L$_V$= 4.35$\pm$1.40 M$_{\odot}$/L$_{V \odot}$ (see Section\,\ref{Mass and ratio} for further discussion on this issue). 
   
\section{Discussion}
\label{discussion}
\subsection{Iron abundance}

The derived iron abundance of Sextans~A GC-1 shows it to be metal-poor but still well above the metallicity floor, $\text{[Fe/H]}=-2.14\pm0.04$. 
Our [Fe/H] determination from the UVES data falls between the two estimates obtained from the OSIRIS spectrum, by \cite{Beasley2019} and the method of \cite{Cabrera-Ziri2022}, respectively (although see Section\,\ref{analysis} regarding some caveats of these models in this metallicity regime). 

Concerning the metallicity of the host galaxy, from the mean V - I colour of the RGB \cite{McConnachie2012} reports $\langle$[Fe/H]$\rangle$ $\sim$ -1.85, while from the colour magnitude diagram (CMD) \cite{Dolphin2003} estimated $\langle$[M/H]$\rangle$ = -1.45$\pm$0.20 (with approximate limits of -1.75 < [M/H] < -1.35). 
Thus, the metallicity range of the metal-poor old populations in Sextans A is extended further towards the low-metallicity tail by GC-1. 
This means that the present stellar populations could be as low as $\text{[Fe/H]}=-2.14\pm0.04$.

According to \cite{Mateo1998}, Sextans~A, WLM, and NGC~6822 have globally similar SF histories as derived from the CMDs. Namely, all three dIrrs had significant SF more than 5-10 Gyr ago, followed by a pause in SF, and then recent SF events within the past Gyr. In \cite{Larsen2022} the iron abundance measured for WLM GC is $\text{[Fe/H]}=-1.85\pm0.03$, which is more metal-rich compared to Sextans~A GC-1; moreover, the iron abundances of three GCs in NGC~6822 were found to be higher than in WLM GC and Sextans~A GC-1. Therefore, Sextans~A GC-1 is the most metal-poor of the studied GCs within these three galaxies that share similar SF histories.

\subsection{$\alpha$ element abundances}
\label{Subsect:mg}

The [$\alpha$/Fe] abundance ratio is susceptible to the ages of stars. This is due to the rapid dispersion of $\alpha$ elements into the ISM by core-collapse SNII from massive stars, occurring on short timescales. In contrast, Fe and Fe-peak elements are dispersed by SNIa, which requires a white dwarf and takes place on longer timescales than SNII \citep{Gilmore1998, Maoz2012}. 
    This means that old stars will be enhanced in [$\alpha$/Fe] with low metallicities, while younger stars will be depleted in the [$\alpha$/Fe] ratio but have higher metallicity. This creates a characteristic ‘knee’ in the [$\alpha$/Fe]–[Fe/H] plane \citep{Tolstoy2009}.
$\alpha$ elements can actually be divided into two groups: those formed during a hydrostatic carbon and neon burning (Mg and O) and those formed in the explosion event of TypeII supernovae (Si, Ca, and Ti) \citep{Woosley1995, Kobayashi2006, Pagel2009book}.

The derived abundances of Ca, Ti, Si, and Mg, and the mean $\alpha$ abundance, are shown in Figure\,\ref{Fig_alpha_elemVSfe}.
In Sextans~A GC-1, both [Ca/Fe] and [Ti/Fe] are moderately enhanced.
Within the uncertainties, the abundances of Si, Ca, and Ti are all consistent with being enhanced relative to a scaled-solar composition and resemble those typically observed for GCs in the Local Group.
In particular, the highlighted GCs in Figure\,\ref{Fig_alpha_elemVSfe} are: N147~PA-1 (red circle) for [Ca/Fe] and [Si/Fe]; Fornax~3 (red star) for [Ca/Fe], [Ti/Fe], and [Si/Fe]; and M31 358-219 (red square) is the closest match for [Ti/Fe] abundance ratios in Sextans~A GC-1. The most occurent GCs that resemble the majority of the element abundances are N147~PA-1 and Fornax~3, hence these two GCs are highlighted in the mean $\alpha$ abundance plot (bottom plot in Figure\,\ref{Fig_alpha_elemVSfe}).

The derived Mg abundance ratio value for Sextans~A GC-1 is extremely depleted, $\text{[Mg/Fe]}=-0.79\pm0.29$. The NLTE corrections for the [Mg/Fe] abundance ratio were computed at [Mg/Fe] = -0.1 because the grid does not extend to such low [Mg/Fe] ratios. However, it does not influence the conclusions since the NLTE corrections are small (see Table\,\ref{tab:abundances}) compared to the depletion. There are no GCs that exhibit such a low value of Mg in the Local Group. That is also not in agreement with the value of Mg abundance in the host galaxy that was measured in \cite{Kaufer2004} to be $\langle$[$\alpha\text{ (Mg)/Fe II, Cr II]$\rangle$}=-0.11\pm0.02\pm0.10$; however, this value was measured for a young population represented by supergiants ($\approx$ 10 Myr).

Strongly depleted [Mg/Fe] ratios of extragalactic GCs are frequently found in IL studies, which still remains puzzling \citep{Colucci2009, Larsen2012, Larsen2014}. 
An internal spread in the [Mg/Fe] abundance ratios has been observed in some MW GCs, with a fraction of the GC stars having lower [Mg/Fe] ratios than field stars of a corresponding metallicity. One example is M13, for which a sample of stars was examined by \cite{Sneden2004}.
The study showed clear Na-O and Mg-Al anti-correlations, with a spread in [Mg/Fe] and [Na/Fe] ratios. In particular, [Mg/Fe] varied from -0.2 to +0.4, while [Na/Fe] varied from -0.3 to +0.6. These variations average to $ \text{[Mg/Fe]} = +0.11$ and $ \text{[Na/Fe]} = +0.21$. 
The behaviour of these values is similar to the ones derived with IL for WLM GC -- $ \text{[Mg/Fe]}=+0.04\pm$0.15 and $ \text{[Na/Fe]}=+0.23\pm$0.15 \citep{Larsen2014}. Another IL study of M13 \citep{Sakari2013} found $ \text{[Mg/Fe]} = +0.14\pm0.10$ and $ \text{[Na/Fe] }= +0.33\pm0.16$. 
An even more extreme example is NGC~2419, where 
[Mg/Fe] ratios of stars extend over from -1 to +1 dex \citep{Mucciarelli2012, Mucciarelli2018}.
While these average [Mg/Fe] values ($\langle$[Mg/Fe]$\rangle$=+0.05$\pm$0.08) are somewhat lower than typical values for field stars, they are not nearly as depleted as the value derived for Sextans A GC-1 from our analysis.
Sometimes the [Mg/Fe] ratios are significantly lower than [Ca/Fe] and [Ti/Fe]. 
The reason for the anomalous [Mg/Fe] ratios in GCs like GC-1 might be linked to a more extreme manifestation of anti-correlations such as those observed in M13, and such internal spreads in the [Mg/Fe] ratio might thus be a common feature within extragalactic GCs. It is however puzzling, then, that no MW GCs exhibit internal Mg abundance variations that are sufficiently large to explain cases like GC-1 (e.g. \citealt{Pancino2017}).
While the environment might be expected to play a role, integrated-light studies of entire galaxies have not uncovered Mg/Fe anomalies as pronounced as those observed in GC-1 (e.g. \citealt{Kuntschner2002, Sanchez-Blazquez2006}). However, some evidence for similar abundance patterns has been observed in ultra-faint satellite companions of the Magellanic Clouds such as Car II, with variations in the high-mass end of the IMF suggested as one possible explanation \citep{Ji2020}.

\begin{figure}
\centering
\includegraphics[width=0.49\textwidth]{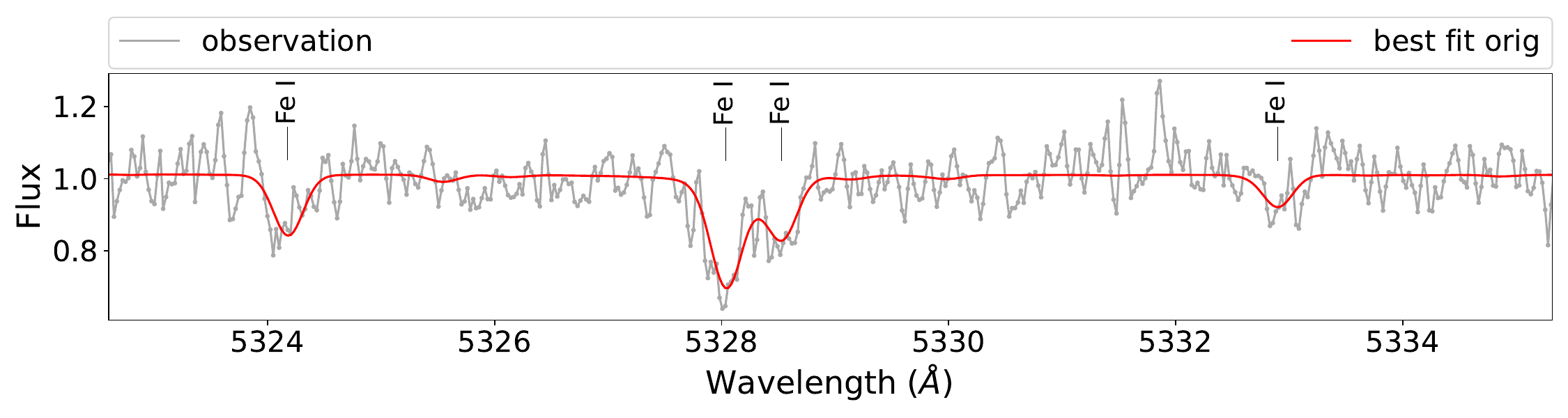}
\includegraphics[width=0.49\textwidth]{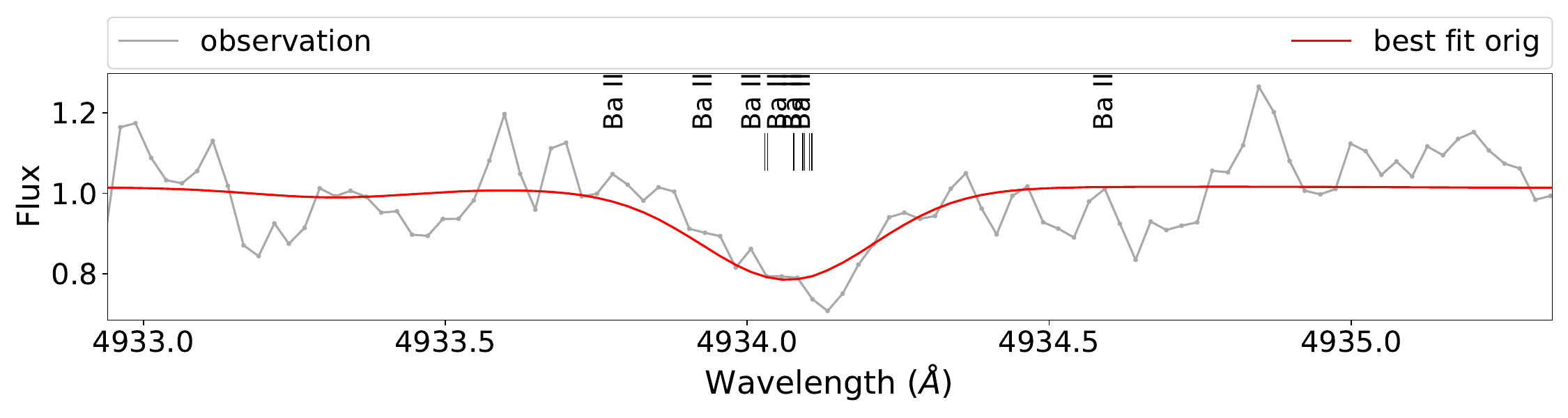}
\includegraphics[width=0.49\textwidth]{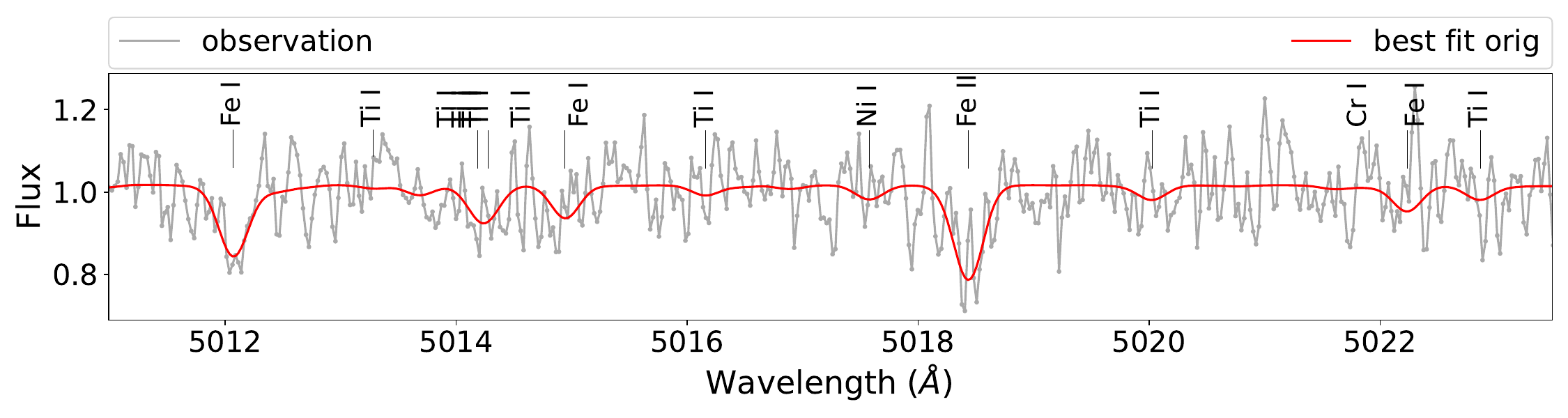}
\caption{Examples of a few spectral windows used. Top: Spectral window at 5328 \AA\  capturing the \ion{Fe}{i} lines. 
Middle: Spectral window of \ion{Ba}{ii} lines at 4934 \AA.
Bottom: Busier spectral region for \ion{Ti}{i} lines at 5014-5016 \AA\ with the presence of other elements such as \ion{Fe}{i} and \ion{Ni}{i}.
The best-fit model is in red and the observed spectrum is in grey.}
\label{spec_wind_fits}%
\end{figure}

\begin{figure}
\centering
\includegraphics[width=0.49\textwidth]{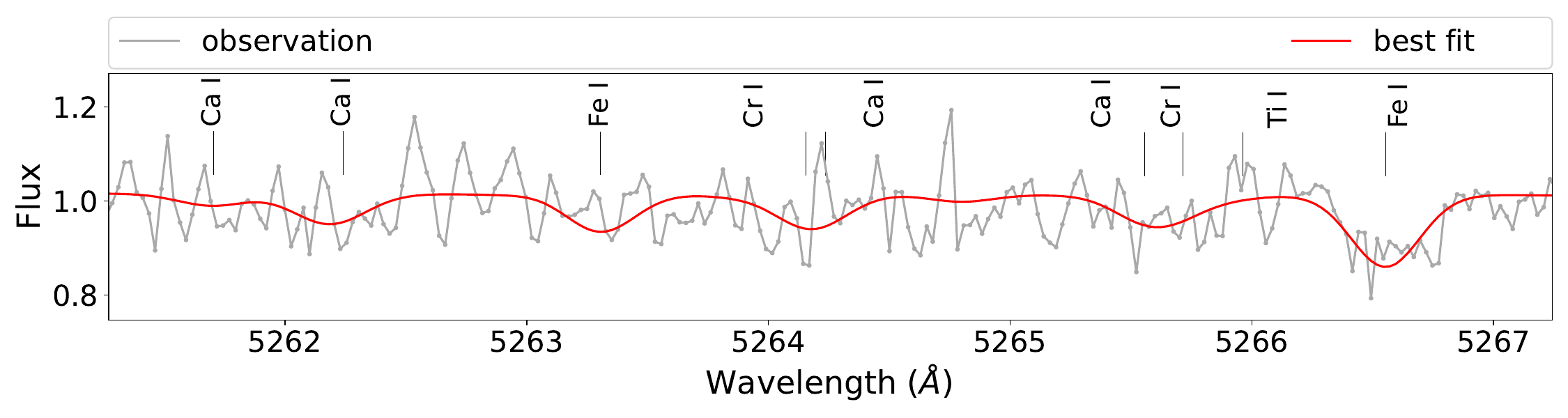}
\includegraphics[width=0.49\textwidth]{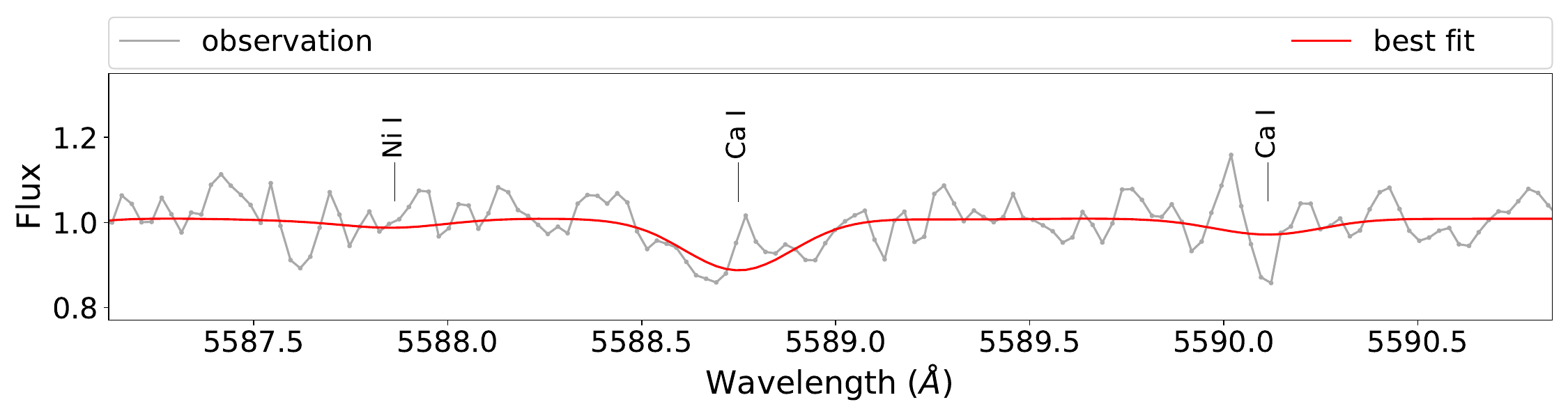}
\caption{Examples of a couple of spectral windows used to derive Ca. Top: Spectral window containing a number of \ion{Ca}{i} lines ranging from 5261.0 -- 5266.0 \AA\ with the presence of other elements such as \ion{Fe}{i}, \ion{Ti}{i}, and \ion{Cr}{i}. 
Bottom: Spectral region of \ion{Ca}{i} lines at 5588.7 \AA\ and 5590.1 \AA  .
The best-fit model is in red and the observed spectrum is in grey.}
\label{Ca_spec_wind_fits}%
\end{figure}

To verify the reliability of the derived value for the [Mg/Fe] abundance ratio, we computed synthetic models for enhanced ([Mg/Fe]=0.3) and solar ([Mg/Fe]=0.0) values for Mg abundance ratio. Figure\,\ref{Fig_Mg_test} shows the resulting difference between the three models: the best-fit model with depleted Mg in blue, solar Mg in red, enhanced Mg abundance in green, and the observed spectrum in grey.
A higher S/N spectrum would lead to a better constraint on [Mg/Fe], but it is already clear from the comparison in Figure\,\ref{Fig_Mg_test} that a favourable model is the depleted Mg model. The reduced $\chi^2$ values derived for these three models are 1.138, 1.177, and 1.185 for the depleted, solar, and enhanced models for the Mg line at 4571 \AA\,, while the reduced $\chi^2$ values for the Mg line at 4703 \AA\ are similarly 1.306, 1.347, and 1.397 for the depleted, solar, and enhanced models.
This test for the used Mg lines at 4571 \AA\ and 4703 \AA\ confirms the Mg deficit in Sextans~A GC-1 and the fact that this low abundance of Mg is a favourable solution. 
Examples of a few other spectral windows and their best fits are given in Figure\,\ref{spec_wind_fits} and Figure\,\ref{Ca_spec_wind_fits}.

The mean $\alpha$-abundance, computed by combining all individual $\alpha$ elements except Mg, 
$\text{[<Si,Ca,Ti>/Fe]}=+0.34\pm0.15$, 
is enhanced at a level that is typical for GCs in the Local Group  \citep{Larsen2022}.
To summarise, in terms of the $\alpha$-element abundance, the Sextans A GC mostly resembles GCs in the Local Group, with the one exception of the strongly depleted [Mg/Fe] value.

\subsection{Fe-peak element abundances} 
The Fe-peak elements were produced through thermonuclear reactions in SNIa and massive stars \citep{Kobayashi2020, Minelli2021}. The derived abundances of Cr, Mn, Sc, and Ni are shown in Figure\,\ref{Fig_other_elemVSfe}.
The [Cr/Fe] abundance ratio of Sextans~A GC-1 is enhanced and it is higher than in the majority of the Local Group GCs. The single cluster that has a higher ratio and is still within the error range is M33 H38, $ \text{[Cr/Fe]} =+0.626\pm$0.161 (red cross in the second plot in Figure\,\ref{Fig_other_elemVSfe}). 

The value of Mn abundance derived is solar and resembles some of the GCs within the Local Group: N147 HIII and PA-1. These GCs also show solar and enhanced abundances (shown in yellow and red circles in Figure\,\ref{Fig_other_elemVSfe}, respectively). This element has a large uncertainty not only for this cluster but for others as well. This element has only two spectral windows fitted, which results in large uncertainty if the two derived values are not in perfect agreement.

The elevated [Sc/Fe] abundance ratio of this GC fits the Local Group picture. 
Sc is also often a tracer of $\alpha$ abundance, which is in agreement with Ca, Ti, and Si, which are scaled-solar or enhanced in Sextans~A GC-1. In the top plot of Figure\,\ref{Fig_other_elemVSfe} the highlighted GCs are N147 PA-1 (red circle) and Fornax 3 (red star), as these two GCs resembled the majority of the $\alpha$ elements and some Fe-peak elements. It is noticeable here that Sextans~A GC-1 and N147 PA-1 are not in perfect agreement for the [Sc/Fe] value; however, this
abundance ratio can be quite diverse among Local Group GCs. It ranges between
solar 0.0 to enhanced 0.4 dex \citep{Larsen2022}. 
Another Fe-peak element that resembles the current pattern seen in the Local Group is Ni. The GCs with the closest [Ni/Fe] are Fornax 3 and M31 358-219 (red star and red square in Figure\,\ref{Fig_other_elemVSfe}, respectively).

\subsection{Heavy element abundances}
Heavy elements derived in this study are Ba, Cu, Zn, and Eu. Ba is mainly created by the slow (s-) neutron-capture process \citep{Burris2000}, while Cu and Zn are thought to form via weak s-processing \citep{Pignatari2010}. Eu is mainly created in a rapid neutron-capture (r-) process \citep{Burris2000}.
The [Ba/Fe] abundance ratio is depleted to a similar value as in N147 HIII (yellow circle in the fourth plot in Figure\,\ref{Fig_other_elemVSfe}).
Figure\,\ref{Fig_range_elemVSfe} shows the abundances for the Cu, Zn, and Eu abundance ratios. These are the elements for which only the upper limit of the value was found. All of the upper limits appear to be consistent with the current picture of the abundance pattern of the Local Group GCs; for example, M31 358-219 (red square in the first plot in Figure\,\ref{Fig_range_elemVSfe}) is particularly close in a value of [Cu/Fe].

\subsection{Mass-to-light ratio and dynamical mass}
\label{Mass and ratio}
The calculated value for the dynamical mass is $M_{dyn}=(5.18 \pm1.62) \times 10^5 M_{\odot}$, 
which is a few times higher than the median mass of GCs in the MW ($\approx 2 \times 10^5 M_{\odot}$, 
e.g. \citealt{Baumgardt2018}). GC-1 is within the ranges found in the Local Group, as significantly more massive GCs with masses well above $10^6$ $M_\odot$ exist in the Local Group (e.g. \citealt{Strader2011}).  

The mass-to-light  ratio of Sextans~A GC-1 was computed to be high: 4.35$\pm$1.40 M$_{\odot}$/L$_{V \odot}$.
The average value of M/L$_V$ in the Local Group is 1.45 M$_{\odot}$/L$_{V \odot}$ \citep{McLaughlin2000}, although higher values have been found for GCs in Cen A, NGC~5128, where the ratio varies between 1.1 - 5 M$_{\odot}$/L$_{V \odot}$ \citep{Rejkuba2008}. The latter study also showed an unexpected trend that some blue and metal-poor GCs have rather high mass-to-light ratios of 4-5 M$_{\odot}$/L$_{V \odot}$, 
in contrast to the expectation from simple stellar population models that the M/L ratio should increase with metallicity.
Similarly, \cite{Strader2011} found that the optical M/L ratios of M31 GCs decline with increasing metallicity.
Given that Sextans A GC-1 is also metal-poor and has a high M/L ratio, it provides another data point that reinforces these trends.

If the dynamical mass is correct -- there is no systematic instrumental influence -- it could mean a couple of different scenarios that would require different ways to prove or refute them.
\cite{Kroupa2013} discussed the possibility of two types of IMF modifications that could enhance the M/L ratio for ultra-compact dwarfs. For a top-heavy IMF, the ratio would be high only if the stellar population were old. This is possible for an old population because the massive stars have become non-luminous remnants.
For a bottom-heavy IMF, instead, the M/L ratio is increased because of an excess of faint main-sequence dwarf stars. 
The two cases are difficult to disentangle by observations, as both populations have low luminosities. 
There are different tracers to identify the type of IMF. 
For a bottom-heavy IMF there should be a characteristic absorption feature of CO index in the spectra from low-mass stars \citep{Mieske2008}. 
If the high M/L ratio is due to a top-heavy IMF then this may be noticeable from low-mass X-ray sources \citep{Kroupa2013}. 
Another explanation for the high M/L ratio could be the presence of an intermediate-mass black hole (IMBH).
An IMBH can be detected through a central rise in the velocity dispersion profile or a shallow central cusp in the surface brightness profile \citep{Brunet2020}.
Another option might be the presence of more exotic non-stellar mass (i.e. dark matter).  

\subsection{Age-metallicity relation}
\label{age Z}

\begin{figure}
\centering
\includegraphics[width=0.49\textwidth]{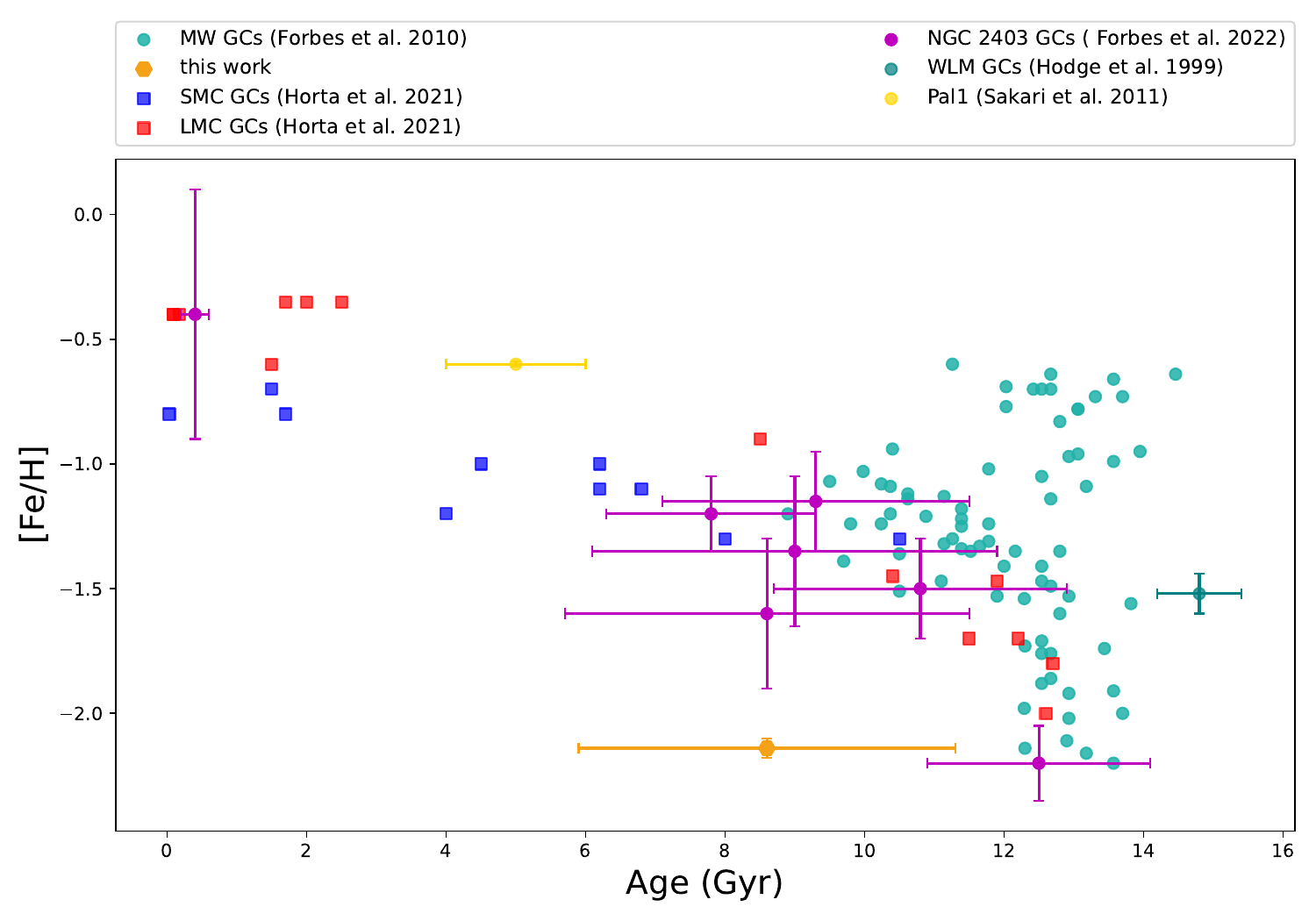}
\caption{Age vs [Fe/H] for GCs in the Local Group and NGC~2403. 
The orange hexagon refers to this work, 
turquoise circles represent GCs in the MW from \cite{Forbes2010},
magenta circles represent NGC~2403 GCs from \cite{Forbes2022},
the teal circle refers to WLM GCs from \cite{Hodge1999},
the yellow circle shows Pal1 from \cite{Sakari2011},
and blue and red squares show GCs with reliable parameters (those with confidence code 1) in the SMC and LMC, respectively \citep{Horta2021}. 
} 
\label{Fig_age_fe}%
\end{figure}

Sometimes this relation is used to define the nature of the MW GC -- they can be recognised as accreted objects as they appear to follow the accretion branch (i.e. the sequence of decreasing age as a function of increasing metallicity) in the bifurcated age-metallicity relation (AMR) of MW GCs \citep{Leaman2013amr, Carretta2022}. The steepness of the AMR is dependent on the mass of the host galaxy; in other words, the very old and metal-rich GCs in the MW are those formed in situ. An example of a clear separation between in situ and accreted MW GCs is shown in \cite{Forbes2010}. This suggests that the detailed age–metallicity distribution of GCs can be used to infer the accretion history of the host galaxy.

Sextans~A GC-1 turns out to be more metal-poor for its age when compared to other GCs in Figure\,\ref{Fig_age_fe}. However, one has to be cautious in drawing conclusions from this given the large uncertainty on the age.
The closest GC system exhibiting a systematically lower iron abundance at a given age when compared to MW GCs is that of the spiral galaxy NGC~2403, which is also isolated.
The two closest to Sextans~A GC-1 are JD1 ($\text{age} =8.6\pm$2.9 and $\text{[Fe/H]} =-1.24\pm$0.31) and C4 ($\text{age} =12.5\pm$1.6 and $\text{[Fe/H]} =-2.07\pm$0.18). The system of GCs in NGC~2403 resembles the SMC system for a leaky-box chemical enrichment model \citep{Leaman2013amr}. Nevertheless, NGC~2403 GCs lie below other GCs of dwarf galaxies accreted onto the MW. This could be a result of the outflow of enriched material \citep{Fraternali2002, Dalcanton2007, Forbes2022}. 
Another interesting target is the WLM GC, since WLM has spent the majority of its lifetime in isolation \citep{Leaman2013}. Unfortunately, there is only a single study with the age determined for this target using the V and I photometry bands obtained with Hubble Space Telescope. Based on that study, WLM GC is one of the oldest (14.8 Gyr) with an iron abundance of $ \text{[Fe/H]} =-1.52\pm$0.08 \citep{Hodge1999}. It should be noted that spectroscopic analyses have found a lower metallicity for the WLM GC \citep{Larsen2022, Colucci2011}.
It could be beneficial to revisit this GC to obtain a more accurate age estimate. 
Hence, while the environment may play a role in setting the AMR for GCs, the WLM GC provides at least one example of a very old, relatively metal-poor GC formed in a low-density environment.

The bottom left side of Figure\,\ref{Fig_age_fe} appears to be underpopulated, which indicates the rarity of young, metal-poor GCs. It would be fascinating to find other isolated GCs to contribute to the same region.

Observational evidence suggests that galaxies in low-density regions tend to form stars with a delay and finish within a longer timescale compared to the galaxies in high-density environments. The SF of field galaxies continues to z$\lesssim$1 and stays at low-level rates even to the present day, while the ones in clusters finish forming stars at z$\gtrsim$2 \citep{Kuntschner2002, Thomas2005}. Additionally, early-type galaxies in low-density environments appear $\sim$1.5 Gyr younger and more metal-rich than in the high-density environments. The favoured scenario is for field galaxies to have a more extended SF history, which results in a less homogeneous stellar population than in the cluster galaxies. Namely, the galaxies in low-density environments are best described by the distribution of the stellar population in which a small portion of young stars is added to an old population \citep{Sanchez-Blazquez2006}. Furthermore, due to the relation found between age and velocity dispersion, authors notice that less-massive galaxies tend to be younger. This seems to be complimentary with the Sextans~A and its GC, as it is a dIrr galaxy with a somewhat younger GC of 8.6$\pm$2.7 Gyr. This galaxy had continuously formed stars in ancient times, which was followed by a break in SF.
Additionally, a small number of stars older than 2.5 Gyr was found, which is again in agreement with \cite{Sanchez-Blazquez2006}. 
This suggests that the SF history of Sextans A is typical of galaxies in low-density environments.

\section{Conclusions}
\label{conclusions}
Sextans~A GC-1 is another cluster that helps populate the low-density environments on the outskirts of the Local Group. It has a curious distribution of $\alpha$ element abundances -- extremely metal-poor Mg, solar-scaled or mildly enhanced Ca and Ti, and enhanced Si. 
In general, this cluster resembles a number of GCs in the Local Group belonging to galaxies such as Fornax and NGC~147. Even though single derived elements resemble the Fornax GCs (Fornax 3 in particular), there is no cluster that would match all of the derived abundances. 
The depletion of Mg to such a low value cannot be explained up to this moment.
Hints of similar (but less extreme) deficiencies of Mg have been observed in the Local Group in GCs such as M13 and NGC~2419 \citep{Sneden2004, Mucciarelli2012}, but none of these clusters have a [Mg/Fe] ratio as low as that observed in Sextans A GC-1.

Sextans~A GC-1 appears to have a high dynamical mass and mass-to-light ratio ($M_{dyn}$=(5.18 $\pm$1.62) $\times$ 10$^5$ M$_{\odot}$ and 4.35$\pm$1.40 M$_{\odot}$/L$_{V \odot}$). This cannot be explained with certainty within this study. A possible solution could include a varying IMF or an IMBH.

This cluster is the first step to start filling up the underpopulated region on the age-metallicity plot. It would be intriguing to find more targets that are metal-poor while being less than 10 Gyr old.

\begin{acknowledgements}

     We thank the anonymous referee for a careful and critical reading of the manuscript. M.A.B. acknowledges financial support from the grant PID2019-107427GB-C32 from the Spanish Ministry of Science, Innovation and Universities (MCIU) and from  the  Severo  Ochoa  Excellence  scheme (SEV-2015-0548). This work was backed in part through the IAC project TRACES which is  supported through the state budget and the regional budget of the Consejería de Economía, Industria, Comercio y Conocimiento of the Canary Islands Autonomous Community. A.G.  thanks Silvia Martocchia for her valuable comments and discussions during this project.
     This study was supported by the Klaus Tschira Foundation.
     
\end{acknowledgements}

%
%

\bibliographystyle{aa}
\bibliography{bibl}

\appendix
\onecolumn

\clearpage
\onecolumn
\section{Chemical abundances }

\begin{longtable}{ccccl}
\caption{Chemical abundances of Sextans~A GC-1.}
\label{tab:spec_win_abun}\\
\hline
\hline
Element & Wavelength range ({\AA}) & LTE abundances & NLTE abundances & SNR per {\AA} \\
\hline
\endfirsthead
\caption{Continued. } \\
\hline
\hline
Element & Wavelength range ({\AA}) & LTE abundances & NLTE abundances & SNR per {\AA} \\

\hline
\endhead
\hline
\endfoot
\hline
\endlastfoot
[Fe/H] & 4573.0 - 4600.0 & - 1.853 $\pm$ 0.103 & - 1.803 $\pm$ 0.103 & 87.6 \\
  & 4600.0 - 4618.0 & - 2.828 $\pm$ 0.302 & - 2.789 $\pm$ 0.302 & 79.3 \\
  & 4631.0 - 4660.0 & - 2.425 $\pm$ 0.152 & - 2.365 $\pm$ 0.152 & 72.3 \\
  & 4705.0 - 4714.0 & - 1.797 $\pm$ 0.169 & - 1.715 $\pm$ 0.169 & 96.4 \\
  & 4724.0 - 4750.0 & - 2.501 $\pm$ 0.156 & - 2.450 $\pm$ 0.156 & 109.9\\
  & 4866.0 - 4883.0 & - 2.580 $\pm$ 0.124 & - 2.561 $\pm$ 0.124 & 89.1\\
  & 4886.0 - 4896.0 & - 2.113 $\pm$ 0.096 & - 2.092 $\pm$ 0.096 & 116.8\\
  & 4897.0 - 4915.0 & - 2.286 $\pm$ 0.165 & - 2.245 $\pm$ 0.165 & 100.3\\
  & 4915.0 - 4929.0 & - 2.181 $\pm$ 0.088 & - 2.157 $\pm$ 0.088 & 103.6 \\
  & 4936.0 - 4944.0 & - 2.939 $\pm$ 0.280 & - 2.905 $\pm$ 0.280 & 95.2\\
  & 4944.0 - 4953.0 & - 2.689 $\pm$ 0.612 & - 2.672 $\pm$ 0.612 & 85.5\\
  & 4963.0 - 4976.0 & - 1.998 $\pm$ 0.143 & - 1.945 $\pm$ 0.143 & 125.4\\
  & 4975.0 - 4998.0 & - 2.577 $\pm$ 0.128 & - 2.527 $\pm$ 0.128 & 87.6\\
  & 5045.0 - 5064.0 & - 1.619 $\pm$ 0.101 & - 1.586 $\pm$ 0.101 & 133.2\\
  & 5066.0 - 5115.0 & - 2.233 $\pm$ 0.061 & - 2.186 $\pm$ 0.061 & 124.6 \\
  & 5118.0 - 5150.0 & - 2.204 $\pm$ 0.060 & - 2.165 $\pm$ 0.060 & 121.8 \\
  & 5250.0 - 5259.0 & - 2.007 $\pm$ 0.148 & - 1.984 $\pm$ 0.148 & 166.6 \\
  & 5271.0 - 5289.0 & - 2.279 $\pm$ 0.109 & - 2.262 $\pm$ 0.109 & 108.1 \\
  & 5300.0 - 5345.0 & - 2.208 $\pm$ 0.058 & - 2.176 $\pm$ 0.058 & 109.0\\
  & 5358.0 - 5375.0 & - 2.379 $\pm$ 0.101 & - 2.337 $\pm$ 0.101 & 119.3\\
  & 5378.0 - 5400.0 & - 1.980 $\pm$ 0.081 & - 1.945 $\pm$ 0.081 & 112.1\\
  & 5400.0 - 5414.0 & - 2.156 $\pm$ 0.084 & - 2.119 $\pm$ 0.084 & 114.6\\
  & 5420.0 - 5460.0 & - 2.279 $\pm$ 0.051 & - 2.245 $\pm$ 0.051 & 99.9\\
  & 5494.0 - 5510.0 & - 1.875 $\pm$ 0.103 & - 1.842 $\pm$ 0.103 & 128.9\\
  & 5529.0 - 5539.0 & - 1.739 $\pm$ 0.210 & - 1.588 $\pm$ 0.210 & 140.1\\
  & 5566.5 - 5590.0 & - 2.031 $\pm$ 0.095 & - 2.028 $\pm$ 0.095 & 116.2\\
  & 5682.0 - 5714.0 & - 2.457 $\pm$ 0.176 & - 2.432 $\pm$ 0.176 & 143.2\\
  & 6053.0 - 6082.0 & - 2.251 $\pm$ 0.164 & - 2.253 $\pm$ 0.164 & 142.7\\
  & 6131.0 - 6140.0 & - 2.216 $\pm$ 0.115 & - 2.216 $\pm$ 0.115 & 155.1\\
  & 6144.0 - 6160.0 & - 2.097 $\pm$ 0.234 & - 2.011 $\pm$ 0.234 & 121.8\\
  
[Ti/Fe] & 4500.0 - 4519.5 & + 0.430 $\pm$ 0.156 & + 0.550 $\pm$ 0.2 & 82.871 \\
  & 4551.0 - 4558.6 & + 0.641 $\pm$ 0.233 & + 0.763 $\pm$ 0.233 & 72.5 \\
  & 4586.5 - 4596.0 & + 0.307 $\pm$ 0.303 & + 0.350 $\pm$ 0.303 & 83.9 \\
  & 4680.0 - 4698.0 & + 0.590 $\pm$ 0.267 & + 0.924 $\pm$ 0.267 & 87.8 \\
  & 4975.0 - 5000.0 & - 0.253 $\pm$ 0.147 & - 0.097 $\pm$ 0.147 & 89.6 \\
  & 5000.0 - 5030.0 & - 0.252 $\pm$ 0.137 & - 0.046 $\pm$ 0.137 & 116.4 \\
  & 5060.0 - 5075.0 & - 0.270 $\pm$ 0.278 & + 0.051 $\pm$ 0.278 & 112.5 \\
  & 5331.0 - 5341.0 & + 0.492 $\pm$ 0.208 & + 0.532 $\pm$ 0.208 & 90.8 \\
  & 5376.0 - 5386.0 & - 0.150 $\pm$ 0.380 & - 0.079 $\pm$ 0.380 & 96.1 \\
  & 5510.0 - 5520.0 & + 0.155 $\pm$ 0.414 & + 0.347 $\pm$ 0.414 & 121.4 \\
  & 5860.0 - 5875.0 & + 0.061 $\pm$ 0.562 & + 0.289 $\pm$ 0.562 & 105.2 \\
  
[Mg/Fe] & 4565.0 - 4576.0 & - 0.622 $\pm$ 0.586 & - 0.619 $\pm$ 0.586 & 63.5 \\
  & 4700.0 - 4707.0 & - 0.837 $\pm$ 0.334 & - 0.849 $\pm$ 0.334 & 91.4 \\

[Si/Fe] & 6150.0 - 6160.0 & + 0.623 $\pm$ 0.256 & + 0.623 $\pm$ 0.256 & 137.9 \\

[Ca/Fe] & 4420.0 - 4440.0 & + 0.212 $\pm$ 0.157 & + 0.211 $\pm$ 0.157 & 89.5 \\
  & 4451.0 - 4461.0 & - 0.604 $\pm$ 0.279 & - 0.616 $\pm$ 0.279 & 80.4\\
  & 4573.0 - 4590.0 & - 0.124 $\pm$ 0.379 & - 0.124 $\pm$ 0.379 & 63.3 \\
  & 5255.0 - 5268.0 & - 0.011 $\pm$ 0.175 & + 0.053 $\pm$ 0.175 & 113.2 \\
  & 5347.0 - 5357.0 & - 0.202 $\pm$ 0.450 & - 0.170 $\pm$ 0.450 & 110.4 \\
  & 5576.0 - 5602.0 & + 0.190 $\pm$ 0.074 & + 0.207 $\pm$ 0.074 & 111.7\\
  & 5852.0 - 5862.0 & - 0.598 $\pm$ 0.456 & - 0.586 $\pm$ 0.456 & 134.7\\
  & 6098.0 - 6127.0 & + 0.349 $\pm$ 0.087 & + 0.318 $\pm$ 0.087 & 133.3 \\
  & 6151.0 - 6174.0 & + 0.003 $\pm$ 0.095 & - 0.036 $\pm$ 0.095 & 146.2\\
  
[Sc/Fe] & 5026.0 - 5036.0 & - 0.287 $\pm$ 0.330 & - 0.287 $\pm$ 0.330 & 113.6\\
  & 5521.0 - 5531.0 & + 0.608 $\pm$ 0.185 & + 0.608 $\pm$ 0.185 & 100.4\\
  & 5638.0 - 5690.0 & + 0.074 $\pm$ 0.148 & + 0.074 $\pm$ 0.148 & 121.2\\
  
[Cr/Fe] & 4611.0 - 4631.0 & + 0.242 $\pm$ 0.148 & + 0.242 $\pm$ 0.148 & 81.7\\
  & 4646.0 - 4657.0 & - 0.866 $\pm$ 0.513 & - 0.866 $\pm$ 0.513 & 76.8\\
  & 4866.0 - 4876.0 & + 0.454 $\pm$ 0.527 & + 0.454 $\pm$ 0.527 & 107.5\\
  & 5063.0 - 5096.0 & + 0.602 $\pm$ 0.362 & + 0.602 $\pm$ 0.362 & 107.15\\
  & 5270.0 - 5281.0 & + 0.895 $\pm$ 0.133 & + 0.895 $\pm$ 0.133 & 65.4\\
  & 5292.0 - 5302.0 & - 0.855 $\pm$ 0.323 & - 0.855 $\pm$ 0.323 & 101.0\\
  & 5341.0 - 5353.0 & + 0.040 $\pm$ 0.145 & + 0.040 $\pm$ 0.145 & 100.1\\
  & 5407.0 - 5413.0 & + 0.226 $\pm$ 0.194 & + 0.226 $\pm$ 0.194 & 128.1\\
  
[Mn/Fe] & 4763.0 - 4790.0 & - 0.879 $\pm$ 0.523 & - 0.694 $\pm$ 0.523 & 91.2\\
  & 6010.0 - 6030.0 & + 0.101 $\pm$ 0.180 & + 0.307 $\pm$ 0.180 & 106.7\\
  
[Ni/Fe] & 4644.0 - 4654.0 & + 0.642 $\pm$ 0.304 & + 0.642 $\pm$ 0.304 & 86.8\\
  & 4681.0 - 4691.0 & + 0.624 $\pm$ 0.366 & + 0.624 $\pm$ 0.366 & 90.1\\
  & 4709.0 - 4719.0 & - 0.145 $\pm$ 0.304 & - 0.053 $\pm$ 0.304 & 89.4\\
  & 4975.0 - 4985.0 & - 0.209 $\pm$ 0.366 & - 0.209 $\pm$ 0.366 & 110.9\\
  & 5075.0 - 5089.0 & - 0.172 $\pm$ 0.168 & + 0.084 $\pm$ 0.168 & 126.6\\
  & 5098.0 - 5108.0 & - 0.055 $\pm$ 0.452 & + 0.179 $\pm$ 0.452 & 110.7\\
  & 5472.0 - 5482.0 & - 0.645 $\pm$ 0.205 & - 0.478 $\pm$ 0.205 & 110.2\\
  & 5707.0 - 5717.0 & - 0.546 $\pm$ 0.491 & - 0.270 $\pm$ 0.491 & 136.7\\
  & 6103.0 - 6113.0 & - 0.087 $\pm$ 0.322 & + 0.169 $\pm$ 0.322 & 152.6\\
  
[Ba/Fe] & 4551.0 - 4558.6 & - 0.860 $\pm$ 0.280 & - 0.878 $\pm$ 0.280 & 72.5\\
  & 4929.0 - 4939.0 & - 0.036 $\pm$ 0.159 & - 0.097 $\pm$ 0.159 & 108.4\\
  & 6135.0 - 6145.0 & - 0.430 $\pm$ 0.207 & - 0.563 $\pm$ 0.207 & 114.1\\

[Cu/Fe] & 5101.0 - 5112.0 & <-0.355 &  & 110.1\\

[Zn/Fe] & 4717.0 - 4727.0 & <+0.143 &  & 96.1\\

[Eu/Fe] & 4431.0 - 4441.0 & <+0.766 &  & 53.5\\

\end{longtable}

\end{document}